# Navigating the nexus: a perspective of centrosome -cytoskeleton interactions

*Subarna Dutta[1,2], Arnab Barua[3]*

1   Department of Biochemistry, University of Calcutta, 35, Ballygunge Circular Road, Kolkata-700019.West Bengal. India.
2   Theomics International Private Limited.28, Income Tax Layout, Sadananda Nagar, NGEF Layout, Bengaluru – 560038, India.
3 Tata Institute of Fundamental Research, Hyderabad - 500046, India

 Correspondence: suborna.dutta@gmail.com

## Abstract:

A structural relationship between the centrosome and cytoskeleton has been recognized for many years. Centrosomes typically reside near the nucleus, establishing and maintaining the nucleus-centrosome axis. This spatial arrangement is critical for determining cell polarity during interphase and ensuring the proper assembly of the spindle apparatus during mitosis. Centrosomes also engage in physical interactions with various components of the cytoskeleton, balancing internal cellular architecture and polarity in a manner specific to tissue type and developmental stage. Numerous crosslinking proteins facilitate these interactions, promoting both cytoskeletal and centrosomal nucleation.
This article provides an overview of how cytoskeletal elements and centrosomes coordinate their actions to regulate complex cellular functions such as cell migration, adhesion, and division. The reciprocal influence between cytoskeletal dynamics and centrosomal positioning underscores their integral roles in cellular organization and function.

## Introduction:

A cell is a highly complex organization responsible for various functions such as migration, growth, and cell-cell communication. To achieve these functions, the cytoskeleton plays a crucial role by providing structural support and acting as the architectural framework of the cell. Unlike static structures, the cytoskeleton is dynamic, constantly assembling and reassembling itself.

The centrosome, discovered in the late 1800s (Schatten et al., 2022), initially posed challenges due to its small size and the limitations of early imaging techniques (Boveri et al., 1901). Early studies using electron microscopy could only analyze centrioles, thereby hindering a comprehensive understanding of the centrosome's overall structure. However, the advent of immunofluorescence microscopy reignited interest in investigating its complexity (Boveri et al., 1901). The centrosome consists of a pair of centrioles embedded in a protein matrix known as pericentriolar material. Positioned near the nucleus, it plays a pivotal role in coordinating cellular activities such as cytoskeletal organization, cell motility, and polarity (Boveri et al., 1901).

In addition to these functions, centrosomes are involved in critical processes such as DNA replication and damage control mechanisms (Theurkauf et al., 2000), heat stress sensing (Doxsey et al., 2016), and participating in immune responses (Hehnly et al., 2021). Their involvement underscores their significance in maintaining cellular homeostasis. Centrosomes interact with the three main components of the cytoskeleton—microtubules, actin, and intermediate filaments—through direct or indirect connections mediated by associated proteins.

Microtubules act as structural pillars within the cell, facilitating cell locomotion, intracellular transport of organelles, key signaling events, and chromosome separation during mitosis. Actin stress fibers provide load-bearing support, while cortical actin contributes to cell shape maintenance and force transmission across the cell membrane. Intermediate filaments, such as vimentin and lamin, play roles in protecting the nucleus, organizing internal cell structures, and supporting nuclear envelope integrity.

For smooth cellular operation, these components must cooperatively interact. The centrosome has emerged as a key regulator in multiple cellular processes, particularly through its interactions with cytoskeletal proteins. However, gaps remain in understanding how its robust architecture influences centrosome-cytoskeleton crosstalk in various cellular processes. Our review aims to highlight this interplay, emphasizing how these components collectively maintain cellular structural integrity, facilitate crucial processes, and enable effective cell migration.

This review is organized into several parts. First, we provide a brief introduction to centrosome structure and describe how its framework and associated proteins maintain interactions with microtubules and actin. Second, we explore how structural modules of actin, microtubules, and intermediate filaments individually contribute to precise cellular processes. Lastly, we focus on variations in this crosstalk from a disease perspective.

## Centrosome structure and function:

The centrosome is a dynamic, membrane-less structure measuring 1-2 µm in size, situated proximal to the nucleus (Doxsey et al., 2001). It consists of two primary components: two perpendicular centrioles and the Pericentriolar Material (PCM) in which the centrioles are embedded (Doxsey et al., 2001). Each centriole exhibits a barrel-shaped structure comprising nine sets of microtubule triplets. Unlike cytoplasmic microtubules, these centriolar microtubules are significantly stabilized through polyglutamylation, a post-translational modification that ensures their long-term stability (Bornens et al., 1998; Bornens et al., 2002).

The two centrioles are categorized as mother and daughter centrioles, differing structurally and functionally. The mother centriole features distal and sub-distal appendages (Schatten et al., 2022). The sub-distal appendages serve as anchoring sites for microtubules, while the distal appendages facilitate centrosome docking to the plasma membrane during cilia formation. Following mitosis, the daughter centriole matures into the mother centriole.

The PCM, the second major component, is a proteinaceous lattice composed of coiled-coil proteins and hosts numerous centrosomal proteins (Kapoor et al., 2017). While the composition and quantity of centrosome proteins within the PCM matrix are well-documented, the extent of its involvement in protein organization remains uncertain (Bloom et al., 2001). PCM proteins play crucial roles in microtubule nucleation, organization, and severing. Table 1 provides a summary of some centriole and PCM-associated proteins along with their respective functions.

Table 1: List of associated proteins of centriole and PCM

| proteins | Location at the centrosome | function |
|---|---|---|
| CEP164 | Distal appendage | Primary cilia formation |
| Centriolin | Subdistal appendage | Cell cycle progression and cytokinesis |
| Ninein | Subdistal appendage | Microtubule anchoring |
| Centrobin | Daughter centriole | Centriole duplication |
| Centrin | Lumen of both centrioles | Centrosome |
| Cenexin | Distal and subdistal appendage | Lumen formation, spindle orientation |
| Pericentrin | PCM | Centrosome and spindle organization |
| Gamma tubulin | PCM | Microtubule nucleation |
| Katanin | PCM | Microtubule severing |

**Centrosome as microtubule nucleating and anchoring organelle:**

The centrosome functions as a pivotal organelle involved in microtubule nucleation, anchoring, and release processes within the cell (Doxsey et al., 2001). Microtubules originating from the centrosome are anchored with their minus-ends tethered at the centrosome and their plus-ends extending into the cytoplasm. These centrosomal microtubules are composed of 13 protofilaments arranged in a tubular structure, originating from the γ-tubulin ring complex (γ-TuRC) located at the centrosome (Doxsey et al., 2001).

Following nucleation, microtubules are anchored at microtubule anchoring sites found at the sub-distal appendage of the mother centriole (Bornens et al., 2000). Due to their dynamic instability and rapid turnover, these microtubules exhibit high dynamism. During cell division, centrosomes organize microtubules into the mitotic spindle, a self-organized structure critical for chromosome segregation. The spindle's microtubules play roles in exerting and maintaining forces necessary for chromosome separation and spindle integrity (Kapoor et al., 2017).

Spindle-associated microtubules can be categorized into kinetochore microtubules (k-fibers), astral microtubules, and interpolar microtubules. K-fibers attach their plus-ends to the kinetochore, while astral microtubules extend from the centrosome towards the cell cortex, aiding in spindle positioning. Interpolar microtubules are densely packed between the spindle poles. Research by Forth et al. has demonstrated that spindle-associated microtubules generate active and passive forces crucial for spindle function (Kapoor et al., 2017).

Centrosomes also play a role in orienting microtubules, which is essential for proper spindle assembly during mitosis and for defining cell polarity during cell migration (Bloom et al., 2001). Wang et al.

(2017) have shown that the centrosome's location within the cell determines the future rear end of the cell, thus influencing cell polarization.

Conversely, studies indicate that microtubules also contribute to centrosome positioning. Experiments by Thery et al. (2016) have demonstrated that during centrosome migration, there is a notable increase in microtubule density around the centrosome, exerting forces that propel its movement. They further indicate that microtubule stabilization promotes centrosome movement. Together, these findings underscore the reciprocal relationship between centrosome function and microtubule organization (Thery et al., 2016).

Microtubule anchoring mechanisms exhibit significant variation among differentiated cells. In non-attached lymphoid cells, most microtubules are typically anchored at the centrosome. Conversely, in adherent epithelial cells, a majority of anchoring proteins are dispersed from the centrosome, suggesting distinct mechanisms of microtubule organization and anchoring in these cells.

### Centrosome as actin nucleating organelle:

In addition to nucleating and organizing microtubules, centrosomes also play a role in nucleating and polymerizing actin microfilaments. The centrosomal protein Pericentriolar material 1 (PCM1) is involved in regulating the Arp2/3 complex and recruiting WASH to the centrosome, thereby facilitating the formation of the centrosomal actin network (Theurkauf et al., 2001). Moreover, centrosomes can polymerize actin through mechanisms that are independent of microtubules, suggesting that centrosomes serve as hubs for signaling molecules that initiate actin polymerization (Doxsey et al., 2001). This dual role highlights the versatility of centrosomes in orchestrating both microtubule and actin dynamics within the cell.

### Role of actin filament network on centrosome:

Actin filaments are fundamental cytoskeletal proteins crucial for regulating cell shape, intracellular transport, and transmitting mechanical forces that enable locomotion and cell division (Holmes et al., 2011; Pollard et al., 2016). Actin monomers polymerize into thin, flexible fibers through a head-to-tail arrangement, forming structures ranging from nanometer-scale units to micrometer-scale fibers (Holmes et al., 2011; Pollard et al., 2016). The efficient nucleation, distribution, and length variation of actin filaments are controlled by various crosslinking molecules and motor proteins.

During cytokinesis, actin assembles into a contractile ring at the cell's equatorial plane, where filament sizes typically range in the hundreds of nanometers (Schroeder et al., 1972). Despite the microtubule organizing center (MTOC), which includes microtubule arrays and the centrosome, the role of actin filaments has been debated and initially considered insignificant. However, compelling evidence supports their involvement in maintaining cellular shape and regulating crucial processes such as cell division. Actin filaments are essential for establishing the spindle's position, division axis, and timing of cytokinesis (Morin et al., 2016; Lew et al., 1999; Bement et al., 2019).

Recent studies highlight that centrosomal branched actin acts as a negative regulator of microtubule nucleation by interfering with the recruitment of γ-Tubulin Ring Complex (γ-TuRC) (Thery et al., 2019; Baum et al., 2019). Elevated levels of branched F-actin correlate with reduced microtubule arrays, emphasizing the existence of actin-microtubule crosstalk at MTOCs like the centrosome (Thery et al., 2019; Grosse et al., 2019). This interdependency is mitigated when actin is in an unbranched state, which does not contribute significantly to microtubule disassembly. However, studies have demonstrated that F-actin structures directly influence the alignment and dynamics of both polymers (Gueroui et al., 2018; Ross et al., 2020).

Additionally, branched actin filaments create a size-dependent diffusion barrier that restricts access to large molecules ($\geq 6.0$ nm in Stokes radius) at centrosomes, regulating centrosomal homeostasis (Lin et al., 2022). It has been proposed that during anaphase, an increase in branched actin filament pools reduces the permeability of centrosomal diffusion barriers (Lin et al., 2022).

Studies by Kita et al. (2019) have identified two distinct populations of F-actin involved in actin-microtubule crosstalk at mitotic spindles and centrosomes. The first population stabilizes normal mitotic spindle morphology and provides anchorage points for astral microtubule plus ends at the cell cortex (Baum et al., 2009; Kramer et al., 2013). This population also assists in positioning the centrosome by guiding astral microtubule tracks (Pellman et al., 2015). The second population comprises fast-growing actin cables that target the spindle apparatus, influencing spindle pole and cell cortex alignment during mitosis (Bement et al., 2019).

In adherent animal cells on highly adhesive substrates, peripheral nuclear and cortical actin fibers maintain spindle parallelism with the substrate, ensuring centrosome alignment at the cell center and nuclei proximity to junctions (Baum et al., 2009). During 3D epithelial morphogenesis, where cell spreading is restricted, peripheral actin contractility diminishes, causing centrosomes to reposition towards cell junctions that mark the initiation of lumen formation (Martin-Belmonte et al., 2012).

These findings collectively underscore the critical role of F-actin in regulating centrosomal functions and maintaining cellular architecture and dynamics during various cellular processes.

### Role of intermediate filament network on centrosome:

Intermediate filaments are essential components of the cytoskeleton, playing a critical role in maintaining cell shape and tissue integrity. These filamentous networks are distributed throughout the cytoplasm and around the nucleus, where they interact with both microtubules and actin filaments. Unlike actin and microtubules, intermediate filaments are characterized by their strong and rope-like structure, providing mechanical support and stability to cells (Koster et al., 2021).

All cells contain intermediate filaments, although the specific protein subunits vary between cell types. Different types of intermediate filaments are associated with specific cell types: neurons predominantly feature neurofilaments, muscle cells contain desmin filaments and epithelial cells are characterized by keratins. Vimentin and lamins are more widely distributed, found in various cell types across different tissues. Vimentin filaments, for example, often co-localize with microtubules and are particularly concentrated near the centrosome, where they form juxtanuclear caps, especially evident when microtubules are depolymerized (Leonova et al., 1995). Stable expression of vimentin leads to organized juxtanuclear structures rather than dispersed cytoplasmic filaments. Vimentin also plays a role in organizing the pericentriolar material, facilitating centrosome-mediated microtubule regrowth and promoting stable acetylated microtubule levels. Loss of vimentin affects centrosome repositioning during processes such as cell polarization and migration, notably impacting wound closure (Patteson et al., 2024).

Phosphorylated keratin 18 accumulates at centrosomes in a cell cycle-dependent manner, contributing to the interaction between mother and daughter centrioles and aiding in microtubule nucleation (Mao et al., 2020). Trichoplein, an associated protein of keratin intermediate filaments, is crucial for anchoring microtubules at the centrosome by recruiting centrosomal proteins to appendages (Inagaki et al., 2011). Depletion of trichoplein impairs microtubule regrowth from the centrosome.

Lamins, as type V nuclear intermediate filament proteins, function during both interphase and mitosis. Their absence in various cell types can lead to irregular distribution of nuclear pore complexes in interphase and defects in spindle orientation in mitotic neural progenitor cells. Lamins contribute to prophase centrosome separation by regulating dynein distribution through the nuclear envelope via the NPC-bound dynein adapter protein BICD2 (Zheng et al., 2015).

In conclusion, intermediate filaments are versatile structures crucial for maintaining cell shape and function through their interactions, both direct and indirect, with the centrosome.

**Existing Mathematical Modeling to explore Centrosome-Cytoskeleton Interactions:**

Mathematical modeling tools are instrumental in comprehending biophysical phenomena by validating theories against experimental datasets and predicting novel outcomes for subsequent laboratory testing, both in vivo and in vitro. Understanding the interaction between the centrosome and cytoskeletal dynamics, primarily regulated through microtubules (MT) and actin filaments, elucidates how the positioning and movement of the centrosome are orchestrated by internal cellular forces.

The positioning and movement of the centrosome within the cell are influenced by the internal forces generated, can be demonstrated as (1) MT-buckling force (from the nucleus) ($F_{MT-BF}$), (2) Dynein pulling (from cortex) ($F_{Dynein-Cortex}$) (3) Dynein pulling (from the nucleus) ($F_{Dynein-Nucleus}$), (4) MT-polymerization force (resisted by cortex) ($F_{Poly.(Cortex)}$), (5) MT-polymerization force (resisted by cell-membrane) ($F_{Poly.(Memb)}$) and (6) MT-polymerization force (resisted by nucleus) ($F_{Poly.(Nucleus)}$) shown in the Fig.(1)[1].

$$\frac{dx_{cs}}{dt} = F_{MT-BF} + F_{Dynein-Cortex} + F_{Dynein-Nucleus} + F_{Poly.(Cortex)} + F_{Poly.(Memb)} + F_{Poly.(Nucleus)}$$

Here the position of the centrosome is defined by $x_{cs}$. These forces are integrated into a computational model capable of predicting the centrosome's trajectories. The study also explores how the nuclear envelope influences centrosome positioning. Two primary forces play critical roles: the cortical pushing force, which propels the centrosome towards the nuclear envelope, and the cortical pulling force, which is notably weaker and typically counteracts the pushing force only at unrealistically high dynein densities. Interestingly, the pushing force is generally stronger when the centrosome is close to the nuclear envelope and exhibits an exponential relationship with increasing distance from the nucleus, akin to the behavior of the Euler buckling force, which sharply diminishes as distance increases. Conversely, the cortical pulling force remains relatively small and changes minimally over the centrosome-nucleus distance, decreasing as the interaction between microtubules (MTs) and the nucleus diminishes with distance. Balancing these forces ensures the centrosome is anchored approximately 1.9 μm from the nuclear envelope in the perinuclear region, achieved through dynamic interactions between MT forces at the cortex and nuclear envelope. This stability allows the centrosome to maintain its position under varying cellular conditions.

Additionally, the positioning dynamics of the centrosome are influenced by the immunological synapse (IS), a structure formed between immune cells (such as T cells or natural killer cells) and their target cells (typically infected or cancerous). The microtubule organizing center (MTOC), which includes the centrosome, plays a crucial role in T cells by relocating to trigger cell lysis. This relocation involves two mechanisms: the cortical sliding mechanism and the MT end-on capture-shrinkage mechanism. Sarker et al. (2019) employed the "search and capture" model to investigate the effectiveness of the dynein-mediated capture-shrinkage mechanism, finding that the time required for MTs to establish end-on attachments with dyneins at the IS cortex varies with T-cell size and IS location. Efficient immune responses rely on factors such as the number of MTs, MT length, and growth velocity, optimizing centrosome positioning to facilitate target search and analysis.

The integrity of the nuclear envelope depends on lamin mesh structures, prompting future research into how the elastic energy of lamins impacts centrosome positioning. In ongoing studies (Dutta et al., 2024, unpublished), lamin aggregates observed in diseased cells alter elastic energy dynamics, potentially influencing centrosome positioning. Exploring these aspects can illuminate critical connections between nuclear envelope integrity, lamin dynamics, and centrosome behavior in cellular processes.

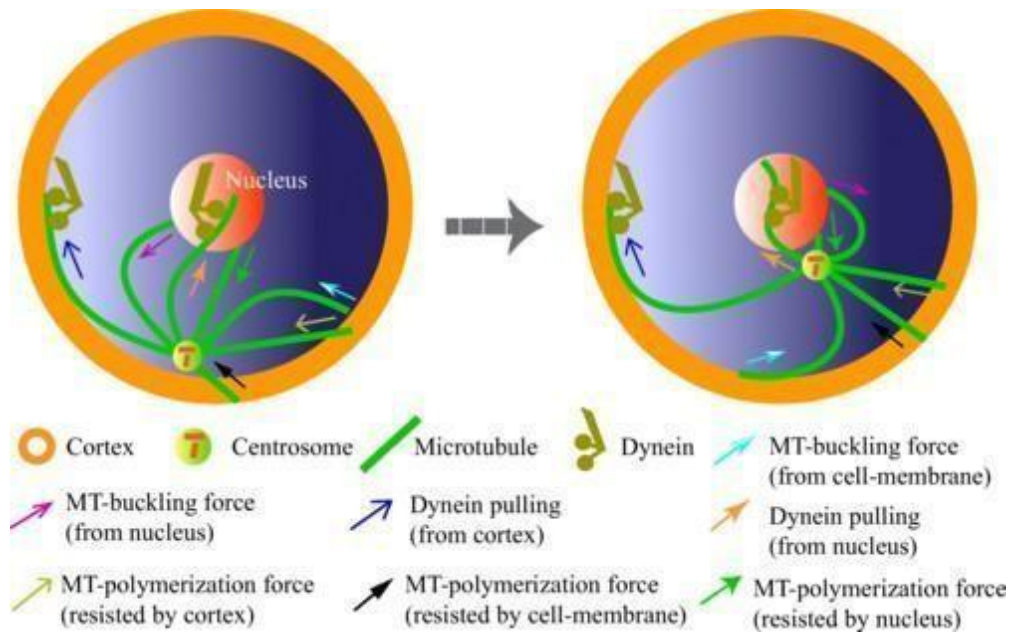

Fig. 1: Schematic diagram of the forces between the centrosome and the various types of cytoskeleton (adapted from Paul et al.2019).

**Centrosome and Cytoskeleton crosstalk with respect to diseases:**
Centrosome abnormalities exert profound effects on both physical and molecular signals, significantly impacting spindle assembly and microtubule-kinetochore attachments. These aberrations are closely linked to chromosomal instability (CIN), contributing to errors in chromosome segregation and replication stress. This cellular stress triggers activation of the cyclic GMP–AMP synthase (cGAS)–stimulator of interferon genes (STING) signaling pathway, thereby influencing the cellular immune response (Liu et al., 2020).
Furthermore, emerging reports highlight the potential direct or indirect roles of the centrosome in aging processes. Aged human oocytes, for instance, exhibit gradual microtubule loss, indicative of structural and functional perturbations in centrosomes. Similarly, in human primary fibroblasts and melanomas, abnormal mitosis and excess centrosomes can lead to cellular senescence after a limited number of divisions. Disruption of core pericentriolar material (PCM) components like PCM-1, Cep192, and NEDD1 further facilitates cell cycle exit, contributing to senescence induction (Hedrick et al., 1999; Merdes et al., 2006).

In cellular contexts such as lymphocyte polarity and neuronal development, centrosomes play crucial roles in organizing the actin cytoskeleton and microtubule arrays (Malbee et al., 2016; Anda et al., 2020). Dysfunctions in centrosomes have been implicated in various brain developmental disorders, including microcephaly primary hereditary (MCPH), Seckel syndrome (SCKL), and microcephalic osteodysplastic primordial dwarfism type II (MOPD II) (Zhou et al., 2021).

In spermatogenesis, proper centrosome function is critical for positioning microtubules and centriole proteins, essential for pronuclear migration. Centrosome aberrations in this process can lead to conditions such as oligozoospermia, asthenozoospermia, teratozoospermia, and acephalic spermatozoa, impacting fertility outcomes (Fishman et al., 2020).

Intermediate filaments, which contribute to mechanical integrity from the nucleus to the cell surface, also play significant roles in cellular function. Imbalances in keratin isoforms, such as K18 and K8, can lead to the formation of intracellular aggregates around centrosomes, compromising microtubule structures and function. This disruption is observed in conditions like alcoholic liver disease, nonalcoholic steatohepatitis, and various hepatocellular neoplasms (Nakayama et al., 2002).

Mutations in lamin A intermediate filaments, such as E385K and N195K, associated with Emery Dreifuss muscular dystrophy and dilated cardiomyopathy, can asymmetrically affect cardiac and skeletal muscle. These mutations alter centrosome orientation and nuclear positioning, disrupting microtubule organizing center (MTOC) dynamics (Schwartz et al., 1999; Folker et al., 2011).

The regulation of different tubulin isoforms and associated microtubule-associated proteins (MAPs) is crucial for maintaining microtubule integrity and function. Mutations in neuron-specific β-tubulin isoform TUBB2A, for example, are linked to conditions like spastic paraplegia and peripheral sensory-motor neuropathy, affecting mitotic spindle dynamics and causing mitotic disruption (Bertini et al., 2020). Similar mutations in TUBB3 and TUBB8 are associated with congenital fibrosis of the extraocular muscles type 3 (CFEOM 3) and developmental abnormalities in human oocytes, affecting spindle morphology and polarization (Wang et al., 2016; Engle et al., 2016).

Overall, disruptions in microtubule regulation and centrosome integrity are implicated in a range of diseases, highlighting their interconnected roles in cellular function and disease pathology.

**Conclusions:**

The cytoskeleton and centrosome are intricately interconnected through direct interactions or via associated proteins and signaling molecules. These interactions undergo various spatiotemporal changes during cell division and migration, influencing processes from spindle organization to cytoskeletal rearrangements and nuclear positioning. Recent advances in animal model organisms and in-vivo cell culture systems have provided insights into the structural and biochemical properties of molecules that mediate this crosstalk.

Cell culture environments differ significantly from native tissue environments. When cells are dissociated from tissues and cultured, they lose their original shape and polarity due to the absence of neighboring cells and the extracellular matrix. Three-dimensional organoid culture systems offer a more physiologically relevant architecture, allowing for detailed investigations into cytoskeleton-centrosome interactions, including centrosome reorientation and cytoskeletal morphology changes.

Technological advancements such as optogenetics, laser dissection, expansion microscopy, and molecular tools like micromanipulators and inducible degrons (e.g., Auxin-inducible degron) enable precise visualization of molecular interactions and modulation of forces during specific phases of mitosis. Cryo-electron tomography has been instrumental in revealing detailed 3D reconstructions of spindle microtubules in mammalian cells (McIntosh et al., 2020; Muller-Reichert et al., 2022).

Synthetic approaches, such as elastic substrates and micropatterning, complement traditional methods by enabling mechanical manipulation of cellular forces. These techniques enhance our understanding of how deformation, curvature, and tension impact cytoskeleton-centrosome crosstalk.

Despite these advancements, several key questions remain unanswered. Recent studies have highlighted the role of actin nucleation around the centrosome during MTOC reorientation in cell division, suggesting the centrosome's pivotal role as a cellular organizing center. Further investigation is needed to elucidate the molecular factors governing this actin nucleation.

Epithelial cells maintain apicobasal polarity through interactions with cell-matrix adhesion proteins like integrins and tight junction proteins. Perturbations in epithelial architecture have been linked to impaired chromosome segregation, which can be mitigated through organoid culture systems dependent on integrin-mediated cell-matrix interactions (Knouse et al., 2018; Amon et al., 2018). This underscores how extracellular contexts influence intrinsic cellular remodeling processes.

In cancer cells, matrix stiffness varies across different cancer types. Synthetic biomimetic environments offer a means to investigate how these variations in cytoskeleton-centrosome structure contribute to aberrant cell division compared to normal cells.

In conclusion, ongoing research using advanced techniques and model systems promises to deepen our understanding of cytoskeleton-centrosome dynamics and their implications for health and disease.

## Author contribution:

SD conceived, designed, and edited the manuscript. SD and AB wrote the manuscript.

## Author Declarations:

The authors have no conflicts to disclose.